\documentclass[twoside,letterpaper]{pazh_mod}

\usepackage{textcomp,mathptmx}
\usepackage{graphicx}

\long\def\***#1{***{\fontfamily{academy}\selectfont #1}***}
\long\def\^^^#1{{\slshape\bfseries #1}}

\begin{document}
\title{Soft X-Ray Sources at the Centers of the Elliptical Galaxies NGC 4472
and NGC 4649}
\author{
  D. A. Soldatenkov\email{sold@hea.iki.rssi.ru}\address{1},
  A. A. Vikhlinin\address{1,2} and
  M. N. Pavlinsky\address{1}
  \addresstext{1}{Space Research Institute, Moscow, Russia}
  \addresstext{2}{Harvard-Smithsonian Center for Astrophysics}
  }
\submitted{\today}

\begin{abstract}
  
  Analysis of recent \emph{Chanrda} observations of the elliptical galaxies
  NGC 4472 and NGC 4649 has revealed faint soft X-ray sources at their
  centers. The sources are located to within $1''$ of the optical centers of
  the galaxies. They are most likely associated with the central
  supermassive black holes. Interest in these and several other similar
  objects stems from the unusually low luminosity of the supermassive black
  holes embedded in dense interstellar medium. Our \emph{Chandra} sources
  have very soft spectra. They are detectable only below $\sim 0.6$~keV and
  have luminosities in the 0.2-2.5~keV energy band of $\sim 6 \times
  10^{37}$~erg~s${}^{-1}$ and $\sim 1.7 \times 10^{38}$ erg s${}^{-1}$ in
  NGC 4649 and NGC 4472, respectively.

\end{abstract}

\section{INTRODUCTION}

Supermassive black holes with masses $\sim 10^{8}-10^{10} M_{\odot}$ are
suspected at the centers of many galaxies on the basis of the stellar
dynamics in central galactic regions (see, e.g. Magorrian et al.1998;
Gebhardt et al.2000; Kormendy and Richstone 1995). Being embedded in a dense
medium, these objects must actively accrete the surrounding material,
resulting in a considerable luminosity of the central galactic
regions. Quasars, blazars, and radio galaxies are different manifestations
of the activity of the central supermassive black holes.

The dilemma known for several years is that, despite the predicted high
accretion rates on the supermassive black holes, the luminosities of the
nuclei of many galaxies are many orders of magnitude lower than those
predicted by standard disk accretion models. For instance, in the elliptical
galaxies NGC 4472 and NGC 4649 discussed below, optical observations
indicate the existence of black holes with masses $\sim 3
\times{10}^{9}{M}_{\odot}$ É $\sim 4\times{10}^{9}{M}_{\odot}$ respectively
(Magorrian et al.1998). The luminosity of the nuclei of these galaxies
calculated using the formula of Bondi (1952) with a 10$\%$ efficiency of
energy release of the total accreted mass and with the parameters of the
interstellar medium derived below is $\sim 5\times{10}^{44}$ erg s${}^{-1}$
for NGC 4472 and $\sim 7\times{10}^{44}$ erg s${}^{-1}$ for NGC 4649.
Actually, however, the luminosities of the nuclei of these galaxies are much
lower (Loewenstein et al.\ 2001).

\begin{figure*}
   \mbox{}\hfill 
   \begin{minipage}[t]{0.98\linewidth}
     \centerline{\includegraphics[width=0.99\linewidth]{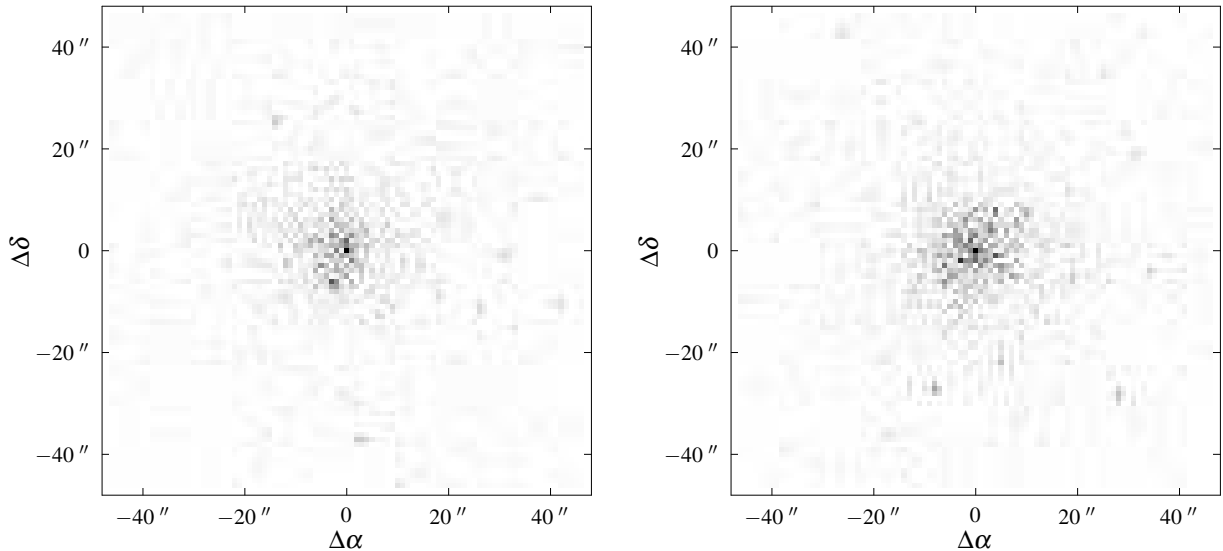}}      
     \caption{Detector images of the galaxies NGC4472 (left) and
              NGC4649(right) in the range 0.2--0.6 keV.
              The image pixel size is $1''$. The images are centered on the
              location of the optical peaks.}
     \label{fig:Tprofile:4874}
   \end{minipage}
\end{figure*}

X-ray observations of NGC4472 and NGC4649 with ROSAT (Irwin and Sarazin
1996; Trinchieri et al.1997) and their optical observations with the Hubble
Space Telescope found no evidence of activity of the central source. Radio
observations (Condon et al 1991; Wrobel 1991) provide the only evidence of
the nuclear activity. They revealed radio sources at the centers of these
galaxies and radio lobes in NGC 4472.

There are theoretical models that explain such low luminosity of the
accreting black holes in galaxies with low-luminosity nuclei: the model with
an advection-dominated accretion flow (Narayan and Yi 1995a,1995b;
Abramowicz et al.1995) and the model with a convection-dominated accretion
flow (Narayan and Yi 2000; Quataert and Gruzinov 2000). The strictest
predictions of these models refer to the radio and X-ray ranges, in which
the theory tested observationally (Di Matteo et al.1999). Therefore, of
considerable interest are observations with \emph{Chandra} whose superior
spatial resolution helps to detect extremely faint point sources. 

In this Paper, we analyze the archival Chandra observations of NGC 4472 and
NGC 4649. We detect previously unreported faint, soft X-ray sources at the
centers of both galaxies. The most plausible interpretation of the nature of
these sources is activity of the central black hole. To our knowledge, this
is the first X-ray detection of the ``quiescent'' supermassive black holes
in elliptical galaxies.

 \begin{figure*}[t]
   \mbox{}\hfill 
   \begin{minipage}[t]{0.99\linewidth}
    \centerline{\includegraphics[width=0.49\linewidth]{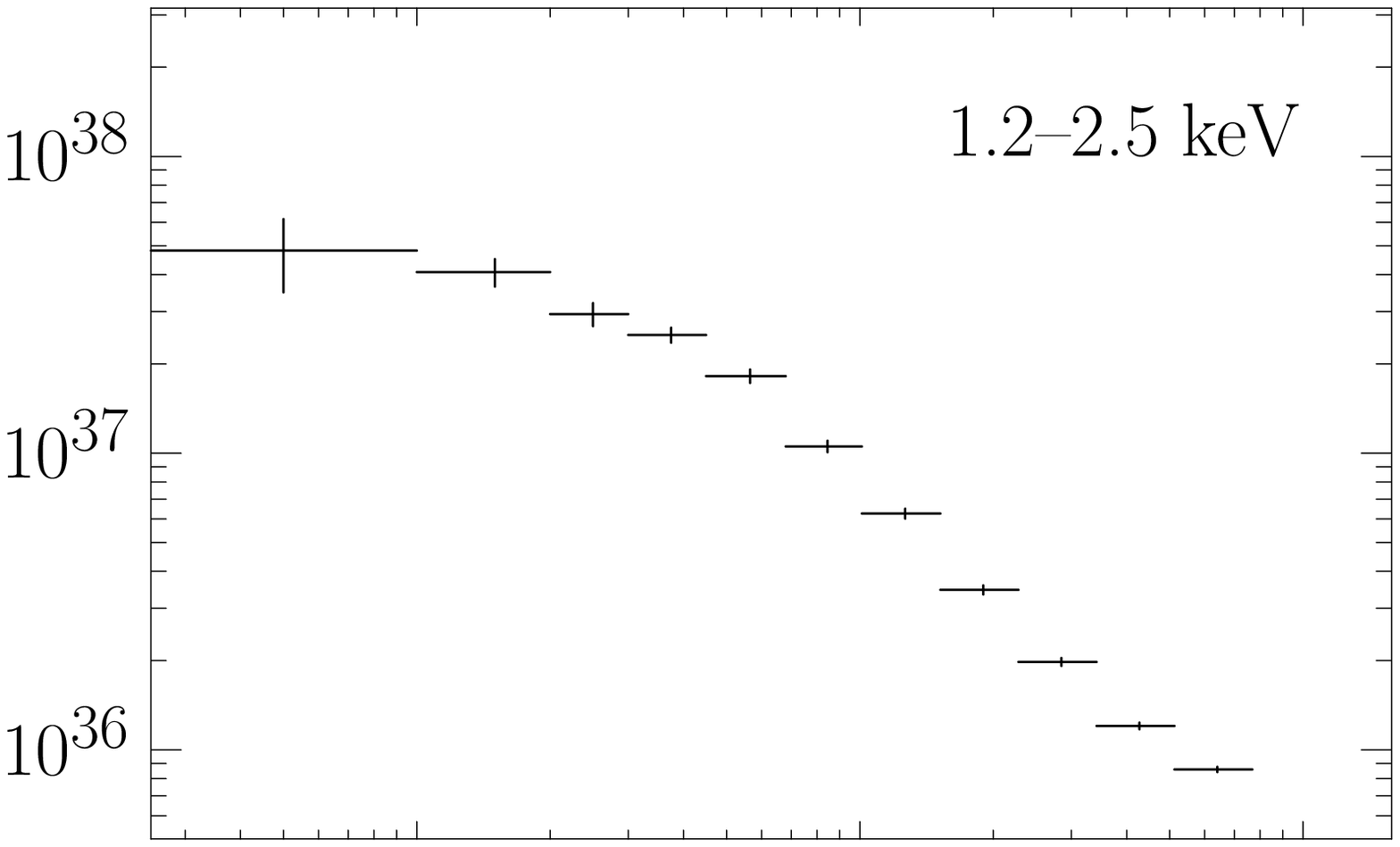}
         \includegraphics[width=0.49\linewidth]{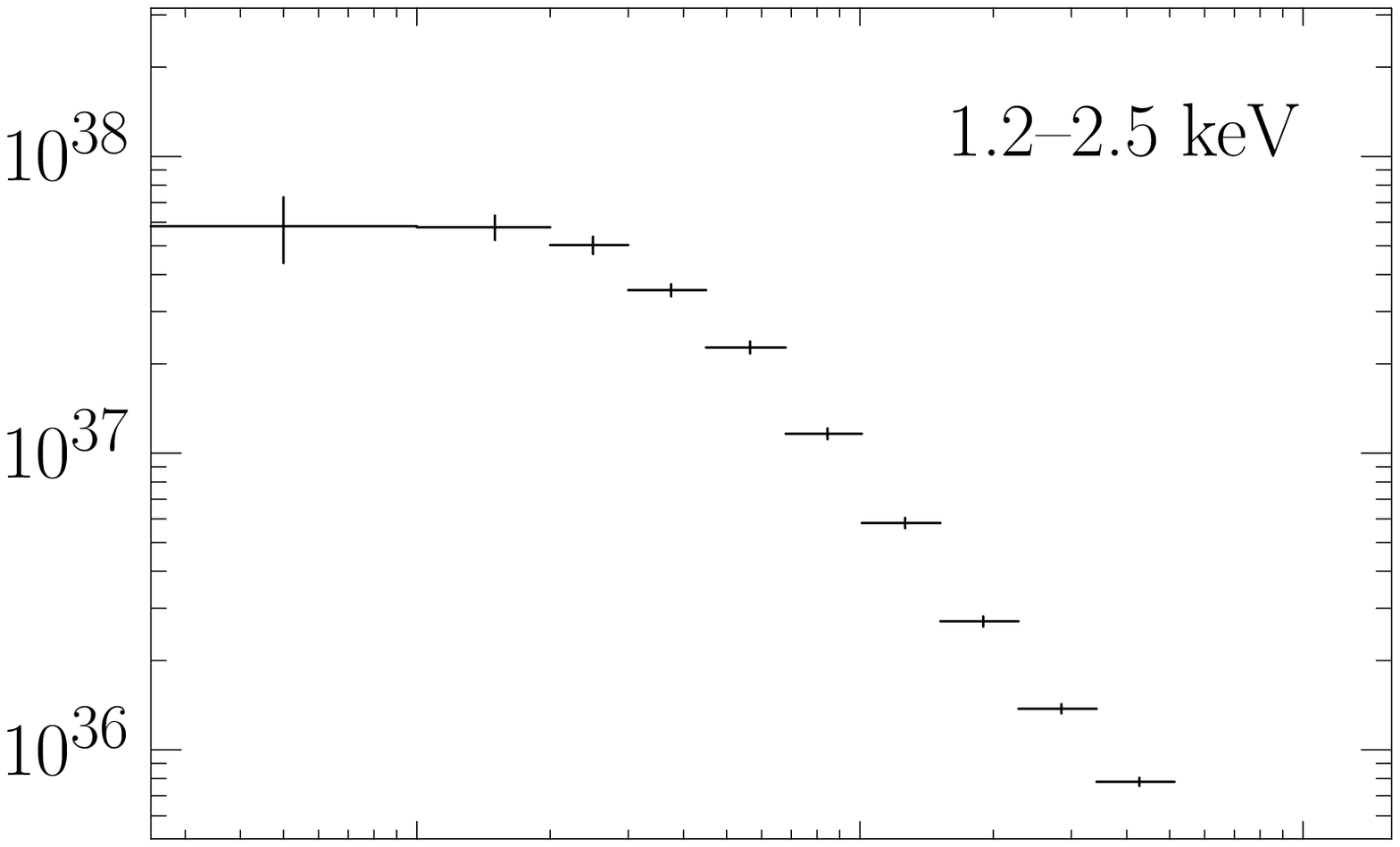}}
    \centerline{\includegraphics[width=0.49\linewidth]{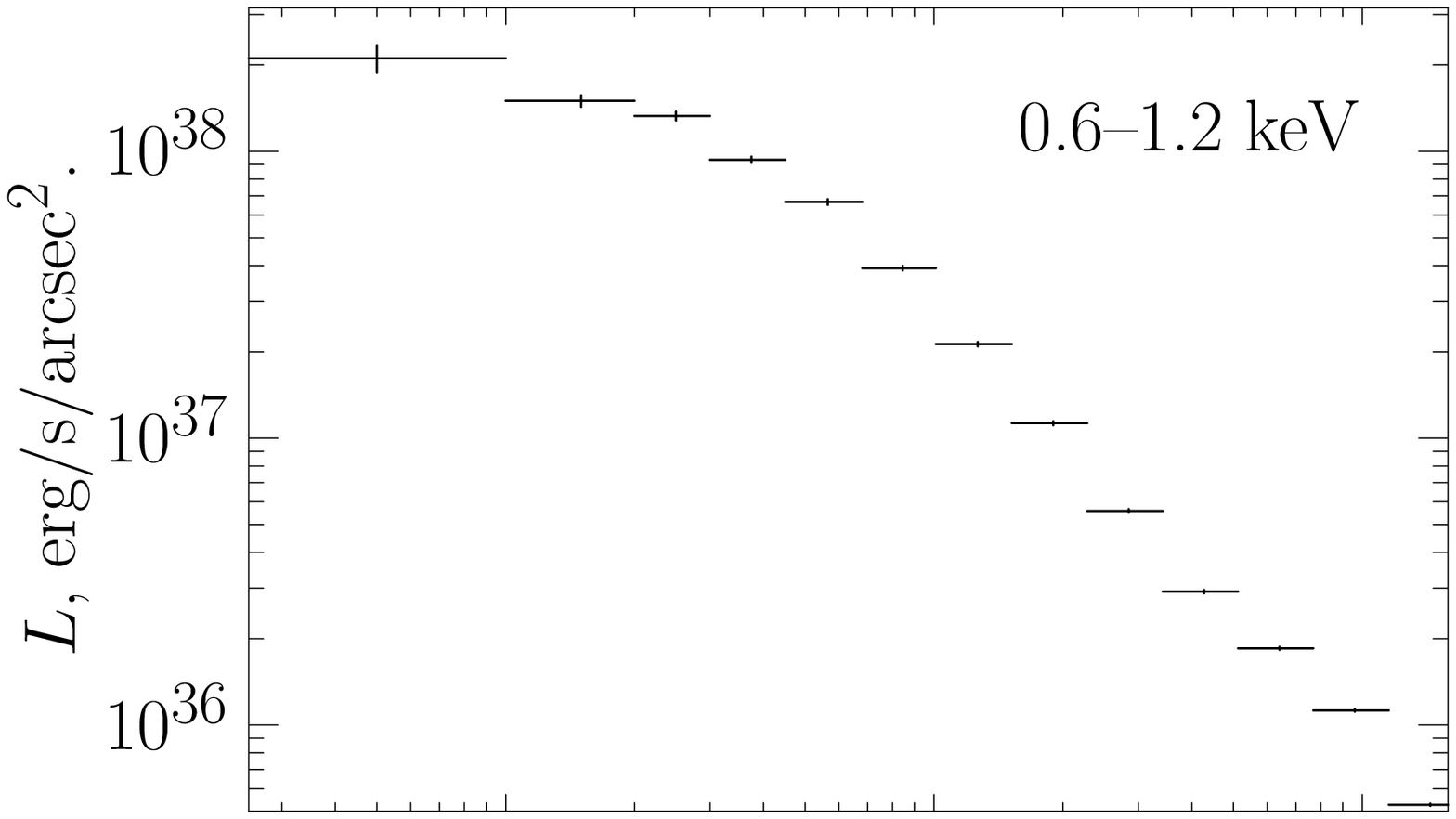}
    \includegraphics[width=0.49\linewidth]{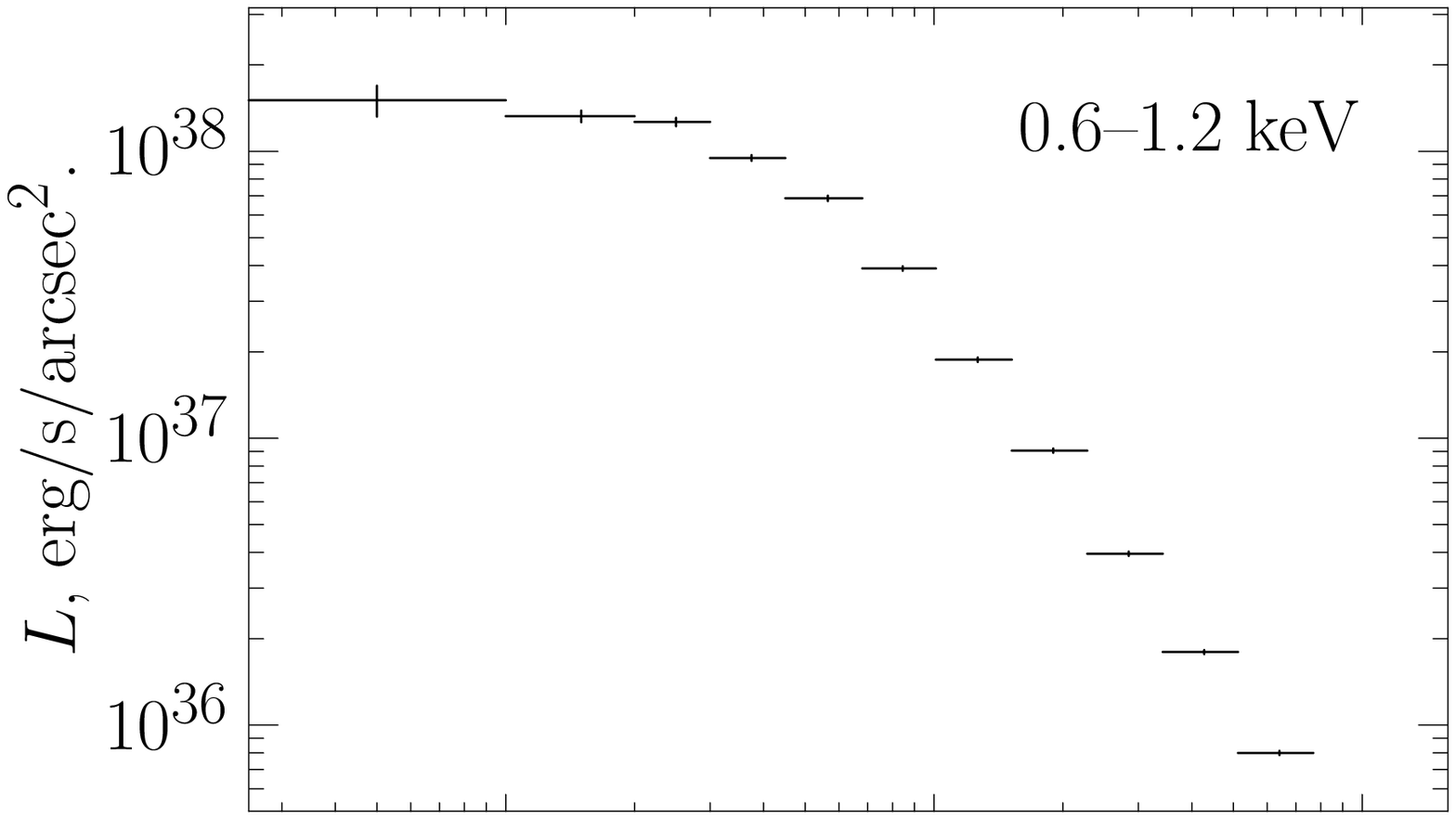}}  
    \centerline{\includegraphics[width=0.49\linewidth]{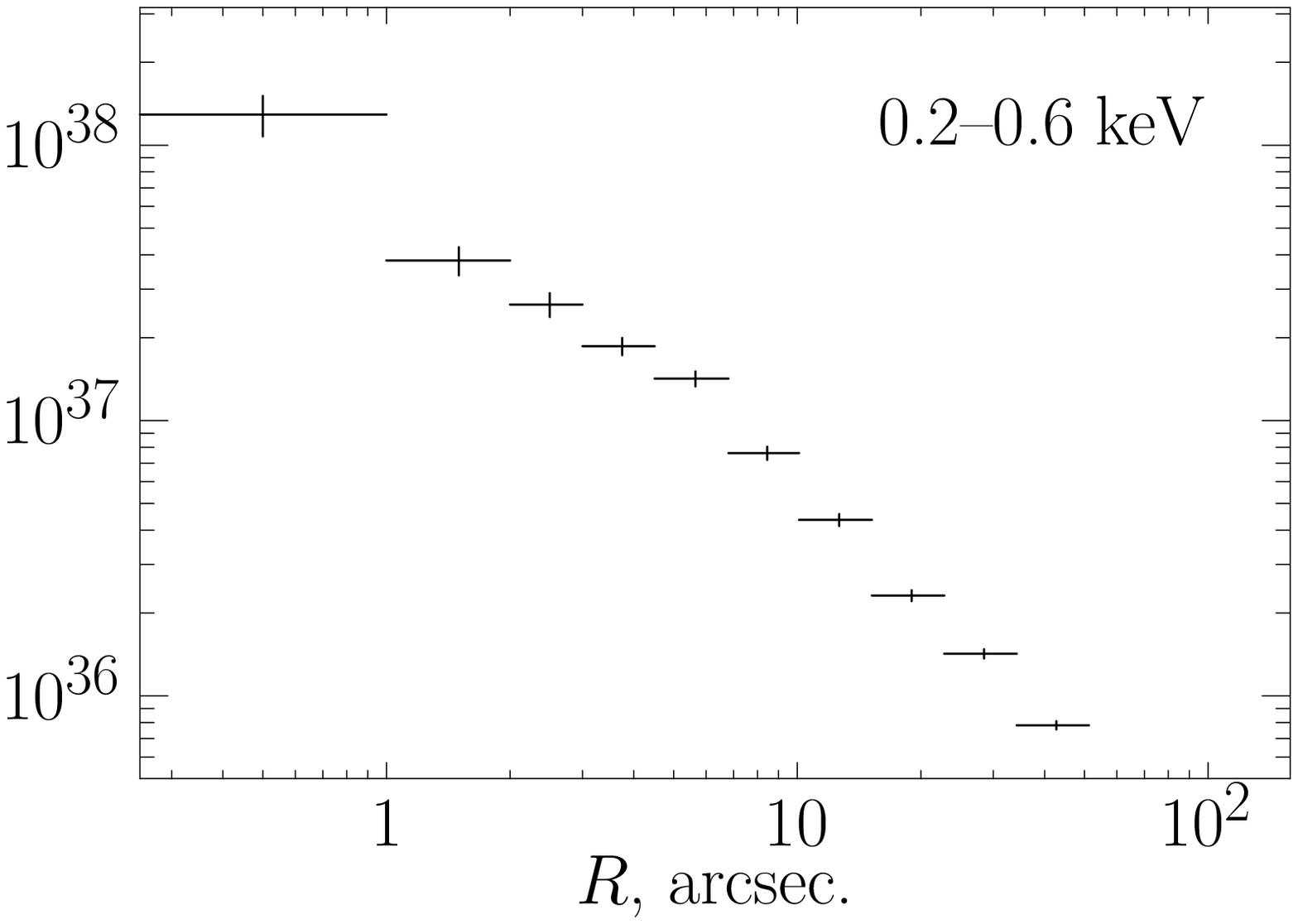}
     \includegraphics[width=0.49\linewidth]{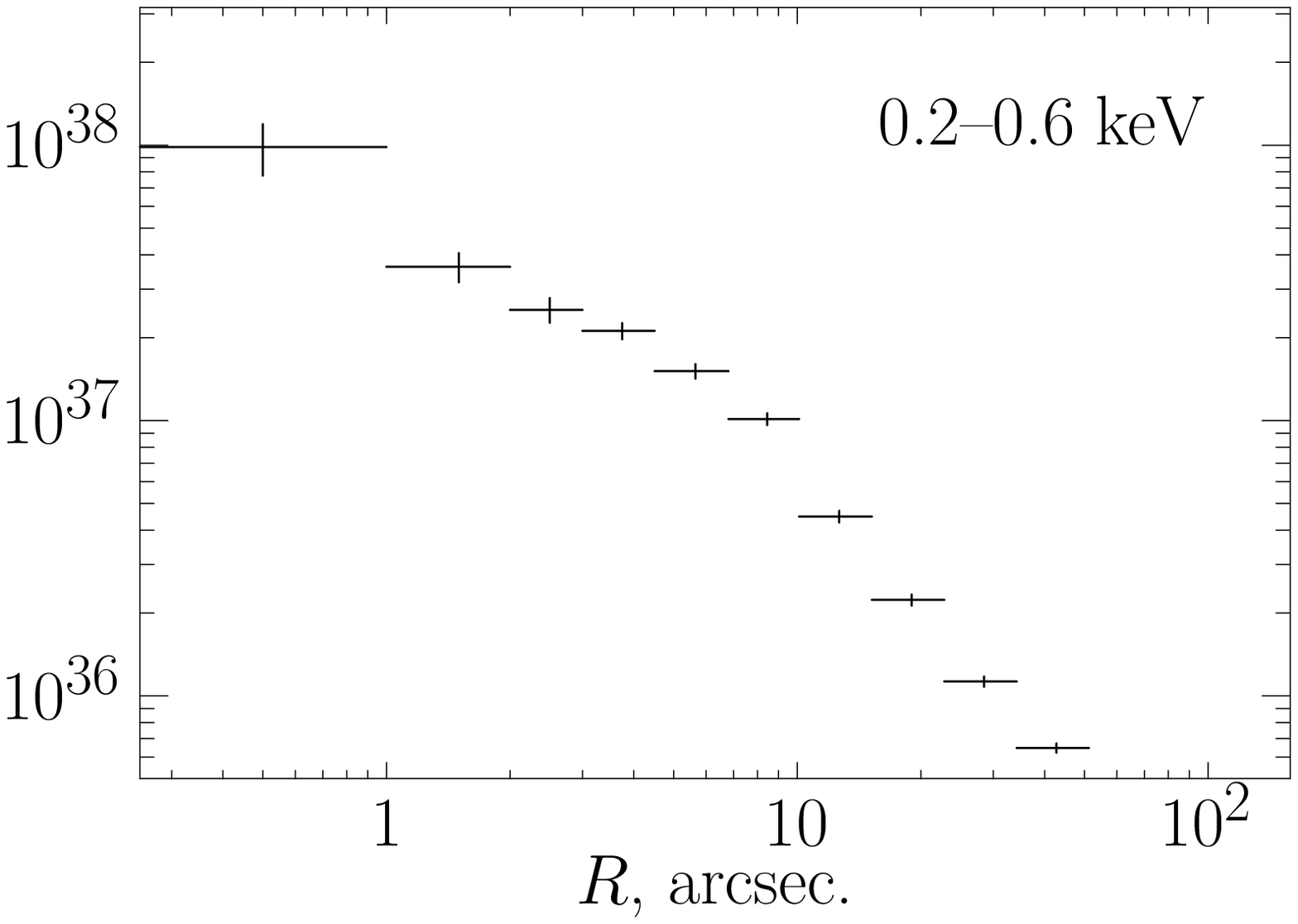}}
    \vspace*{1.5\baselineskip}
    \caption{Surface brightness profiles for the galaxies NGC 4472 (left)
           and NGC 4649 (right) in the energy ranges 0.2--0.6, 0.6--1.2
           and 1.2--2.5 keV.}
    \label{fig:Tprofile:4874}
  \end{minipage}
\end{figure*}

\section{DATA ANALYSIS}

We used the \emph{Chandra} observation of NGC 4472 carried out on June 12,
2000, and of NGC 4649 on April 20, 2000, with a total exposure of $\sim$ 40
000 s each. The ACIS S3 detector was used in both observations. The data
were processed with the standard event filtering criteria and latest
calibration data. We also corrected the ACIS S3 quantum efficiency for the
contamination buildup on the optical blocking filter, which is particularly
pronounced at energies below
1keV.\footnote{asc.harvard.edu/cal/Links/Acis/acis/Cal\_prods/qeDeg/index.html}
During the observations there were flares of the particle-induced background
with a duration of $\sim$ 7000 s for NGC 4472 and $\sim$ 8000 s for NGC
4649. Since we are primarily interested in a small region in the central,
bright part of the galaxies, we opted to keep the data from the flaring
periods.

At low energies, the detector angular resolution degrades from $0.5''$ to
$\sim1''$. Since one pixel on the ACIS S3 detector corresponds to $0.5''$
faint soft X-ray sources are most noticeable when the image is blocked into
$1''$ pixels. The subsequent analysis was carried out on such images.

The X-ray emission from elliptical galaxies is mainly due to the thermal
radiation from hot gas, with a smaller contribution from LMXBs. In the
galaxies NGC 4472 and NGC 4649, the extended emission is symmetric relative
to the galactic center (Fig.1) and no obvious manifestations of the galactic
nuclear activity are observed, with the exception of a small displacement of
the gas centroid in NGC 4472 to the north from the optical center of the
galaxy.

Our subsequent analysis is performed in the energy ranges 0.2--0.6,
0.6--1.2, 1.2--2.5 and 2.5--10 keV. These energy bands approximately bracket
the different emission mechanisms in the interstellar gas. In the first and
third bands, bremsstrahlung and free---bound transitions mainly contribute
to the gas radiation. In the second range, bound---bound transitions on ions
of heavy elements mainly contribute to the spectrum.  The bulk of the
interstellar gas radiation at a temperature $T\sim1$ keV is in the 0.6--1.2
keV band. In the fourth range, 2.5--10 keV, there is virtually no
interstellar gas radiation.

\begin{figure*}[t]
   \mbox{}\hfill 
   \begin{minipage}[t]{0.99\linewidth}
    \centerline{\includegraphics[width=0.49\linewidth]{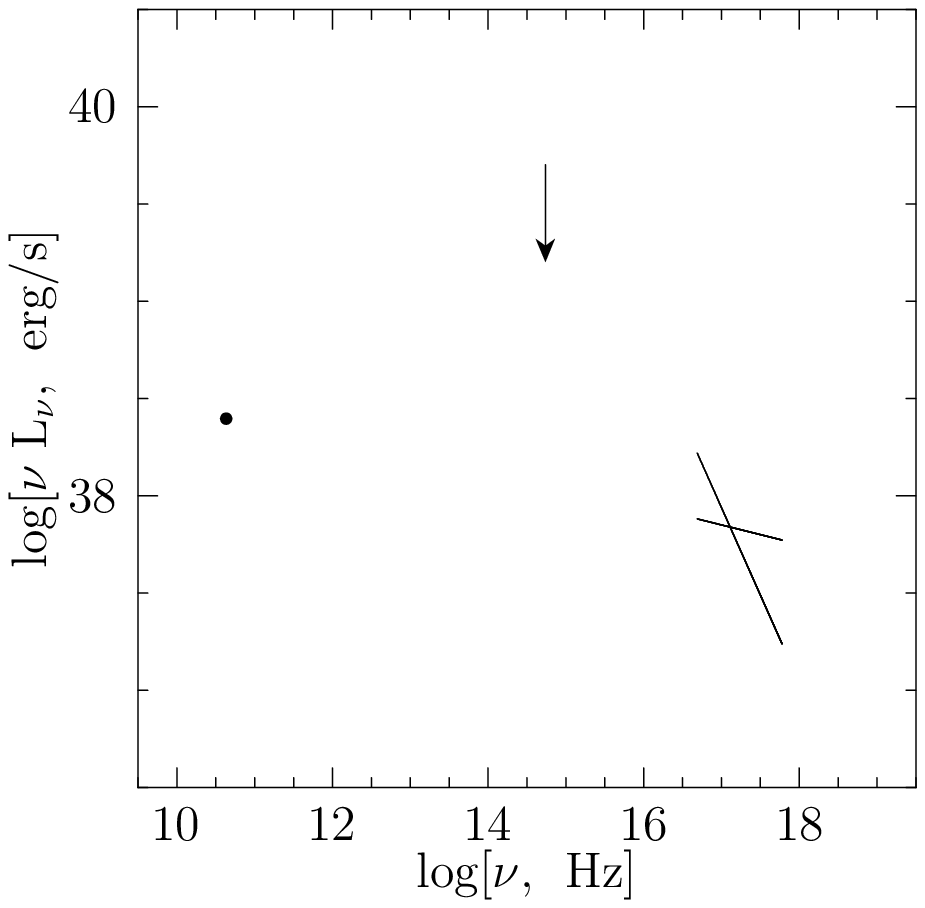}
         \includegraphics[width=0.49\linewidth]{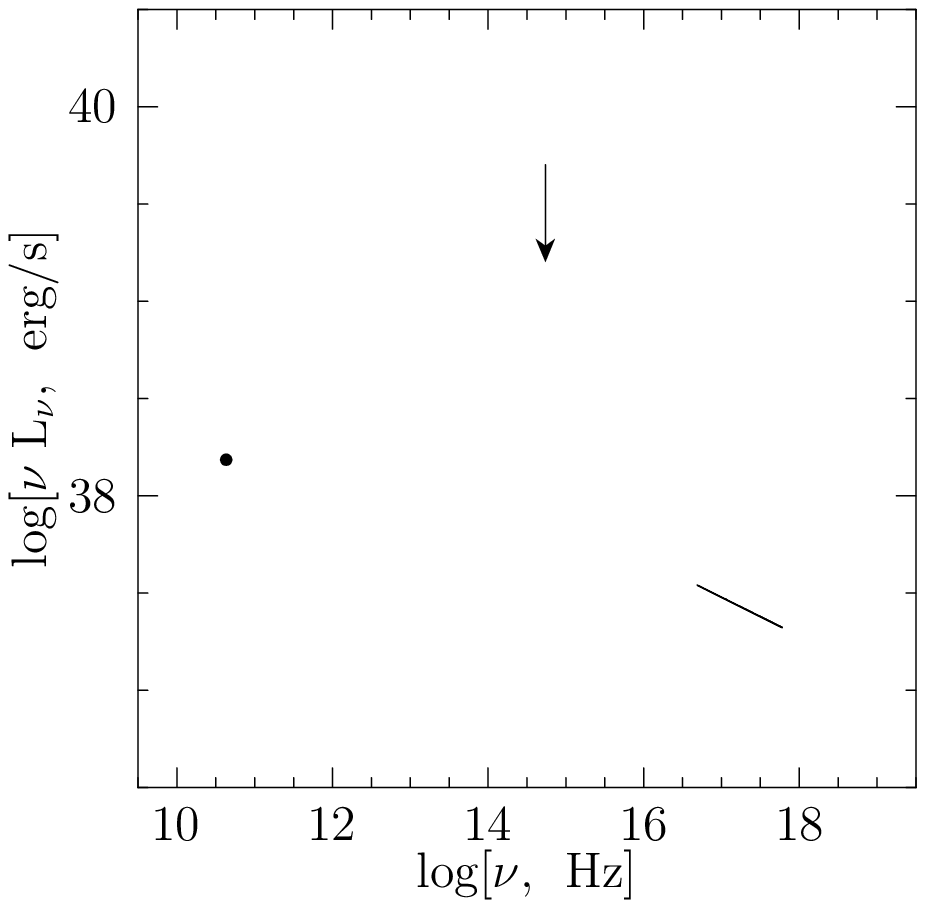}}
     \caption{Radio (dots) and X-ray (lines) fluxes from the central sources
        in the galaxies NGC4472 (left) and NGC4649(right) and upper limits on the
        optical luminosity (arrows)}
    \label{fig:Tprofile:4874}
  \end{minipage}
\end{figure*}

Inspection of the images reveal faint, soft (detectable only in the 0.2--0.6
keV band) X-ray sources near the centers of both galaxies (Fig.~1-2). Their
angular sizes are consistent with the \emph{Chandra} PSF at these energies,
$\sim 1''$. In the 0.6--1.2 and 1.2--2.5 keV energy bands, the compact
emission from the center is undetectable and the surface brightness profiles
are well described by the $\beta$-model (Cavaliere \& Fusco-Femiano 1976):
\begin{equation}
I(r)=\frac{I_{0}}{\displaystyle \left(1+r^{2}/a^{2}\right)^{3 \beta -0.5}}
\end{equation}
with a $\sim 5''$ core. 

The most natural explanation of the central sources is the emission from
from the central supermassive black hole. We also considered, however,
alternative explanations. For example, there may be power-law density peaks
in the ISM distribution and the observed sources is in fact a thermal
emission from the ISM. In this case, the absence of a central peak in the
0.6--1.2 keV band which is dominated by the line emission could be explained
by the resonant scattering (Gilfanov, Sunyaev \& Churazov 1987). However,
this assumption can be rejected, because no central peak is observed in the
1.2--2.5 keV brightness profile, where the contribution from the line
emission is small.

To verify the absence of any systematic offset between the optical and X-ray
positions, we used other point X-ray source in the field. In general, point
X-ray sources in elliptical galaxies are LMXBs, a considerable fraction of
which are located in globular clusters. For instance, in one of the galaxies
studied, NGC 4472, about 40$\%$ of the point X-ray sources are associated
with globular clusters (Kundu et al. 2002). Comparison of the optical HST
images and X-ray source locations shows the validity of the X-ray aspect
solution to within $1''$. Our central X-ray sources coincide with the
optical galaxy centroids to the same accuracy. This is the clearest
indication for association of the detected sources with the galaxy nuclei.

\section{THE SOURCE LUMINOSITIES}

The observed flux from the central sources themselves is low. In the
0.2--0.6 keV band, $22^{-6}_{+7}$ events were detected for the NGC 4472
source and $11^{-5}_{+6}$ events for NGC 4649 (68\% confidence interval
estimated following Gehrels 1986). In both cases, the statistical
significance of the detection is greater than 3$\sigma$. In the 0.6--2.5 keV
band, the flux is statistically significant at a 3$\sigma$ level only for
NGC4472 and is $29^{-11}_{+12}$ photons. In the 2.5--10 keV band the total
flux from the entire central $1''$ region in the galaxies is only a few
photons.

A low flux from the sources does not allow an accurate judgment about their
spectra to be made. Therefore, we used the power-law model to convert
observed counts to the source luminosities. The results are presented in
Table~1; in our calculations, the distance to the two galaxies was assumed
to be 15.3 Mpc (Faber et al. 1997). If the ratio of the 0.6--2.5 and
0.2--0.6-keV fluxes is described by an effective power-law slope, then we
obtain the following photon indices: $2.5\pm0.4$ for NGC 4472 and $> 2.2$ for
NGC 4649. Figure 3 shows the power-law models with the derived parameters of
the photon index and, for comparison, the source radio luminosities at a
frequency of $4.3\times10^{10}$ Hz and the upper limits on the optical flux
at $5.45\times10^{14}$ Hz taken from Di Matteo et al.(1999).

\begin{table}[t]
\begin{center}
\caption{Luminosities of the central sources}\label{tab:profiles}
\def\arraystretch{1.8}
\setlength{\tabcolsep}{1.5pt}
\footnotesize
\begin{tabular}{p{1.5cm}cccc}
\multicolumn{1}{c}{Object} &
\multicolumn{1}{c}{$L_{0.2-0.6}$,  10${}^{37}$~erg/s} & 
\multicolumn{1}{c}{$L_{0.6-2.5 }$,  10${}^{37}$~erg/s} & 
\multicolumn{1}{c}{$L_{2.5-10.0}$, 10${}^{37}$~erg/s} \\ 
\hline
 NGC4472 \dotfill & $9.2^{+2.8}_{-2.4}$ & $7.3^{+2.9}_{-2.6}$ & $< 3.4$  \\
 NGC4649 \dotfill & $6.0^{+3.2}_{-2.6}$ & $ < 3.4$ & $4.6^{+4.7}_{-3.1}$ 
\end{tabular}
\medskip
\begin{minipage}{0.99\linewidth}
\end{minipage}
\end{center}
\end{table}

In conclusion, we can give the interstellar-gas parameters required to
estimate the accretion rate. The gas density at the galactic centers can be
determined by using model (1) for the galactic-gas surface brightness and by
describing its spectrum by the mekal model of radiation from an optically
thin plasma (Mewe et al. 1985; Kaastra 1992; Liedahl et al. 1995) in the
XSPEC code [for details see Voevodkin et al. 2002)]. Our calculations yield
the gas densities $\rho_{g}=(1.0 \pm 0.1) \times 10^{-24}$ g cm$^{-3}$ at
the center of NGC 4472 and $\rho_{g}=(1.1 \pm 0.1) \times 10^{-24}$ g
cm$^{-3}$ at the center of NGC 4649. The gas temperature within $2''$ is
$0.66 \pm 0.02$ keV in NGC 4472 and $0.86\pm0.02$ keV in NGC 4649.

\section{CONCLUSIONS}

We have analyzed the data on the elliptical galaxies NGC 4649 and NGC 4472
that have recently been obtained with Chandra and detected faint, soft X-ray
sources at their centers. The most plausible interpretation of the
luminosities of these sources is activity of a supermassive black
holes. Note that the spectra of the central sources are considerably softer
than those for most of the other objects in the field. However, we cannot
completely rule out the interpretation of the observed central sources as
being the radiation from binary systems of stellar mass.

If our sources indeed represent the emission from the central black holes,
our results is the first X-ray detection of such systems in the quiescent
state. In this case, the X-ray data on these galaxies should be useful for
testing theories that explain the deficit of luminosity in the nuclei of
many galaxies.

\section{ACKNOWLEDGMENTS}

This work was supported by the Russian Foundation for Basic Research
(project no. 00-02-17124) and by the ``Young Scientist'' program of the
Russian Academy of Sciences.

\def\bibname{REFERENCES}

\end{document}